\documentclass[12pt]{iopart}
\usepackage{graphicx}
\begin{document}
\title{Finite axionic electrodynamics from a new noncommutative approach}
\author{Patricio Gaete\dag\footnote{e-mail address: patricio.gaete@usm.cl},
Euro Spallucci\ddag\footnote{e-mail address: spallucci@ts.infn.it}}
\address{\dag\ Departamento de F\'{\i}sica and Centro Cient\'{i}fico-Tecnol\'ogico de Valpara\'{i}so,
 Universidad T\'ecnica Federico Santa Mar\'{\i}a, Valpara\'{\i}so, Chile}
\address{\ddag\ Dipartimento di Fisica Teorica, Universit\`a di \
Trieste and INFN, Sezione di Trieste, Italy}

\begin{abstract}
Using the gauge-invariant but path-dependent variables formalism,
we compute the static quantum potential for noncommutative axionic
electrodynamics (or axionic electrodynamics in the presence of a minimal
length). Accordingly, we obtain an ultraviolet finite static potential
which is the sum of a Yukawa-type and a linear potential, leading to
the confinement of static charges. Interestingly, it should be noted
that this calculation involves no $\theta$ expansion at all. The present
result makes manifest the key role played by the new quantum of length
in our analysis.
\end{abstract}
\submitted
\maketitle

\section{Introduction}

The formulation and physical consequences of extensions of the standard
formalism of field theory to allow non-commuting position operators have
been the object of intensive investigations by many authors
\cite{Witten:1985cc,Seiberg:1999vs,Douglas:2001ba,Szabo:2001kg,Gomis:2000sp,Bichl:2001nf}.
Let us recall here that these new commutation relations were originally
suggested with the goal of avoiding ultraviolet divergences, which appear
within the perturbative approach of quantum field theory \cite{Snyder}.
In this perspective, it should be recalled that recently considerable
attention has been paid to the study of noncommutative field theories due
to its natural emergence in string theory \cite{Witten:1985cc,Seiberg:1999vs}.
In addition to the string interest, noncommutative quantum field theories have also attracted considerable 
attention because some surprising consequences, for example: the ultraviolet-infrared mixing
\cite{Minwalla:1999px}, loss of unitarity in models where time does not commute with space coordinates 
\cite{Gomis:2000zz}, and Lorentz symmetry breaking \cite{Bahns:2002vm}.

It is worth recalling at this point that these studies have been achieved
by using a star product (Moyal product).
More recently, a novel way to formulate noncommutative quantum field
theory (or quantum field theory in the presence of a minimal length) has
been proposed in \cite{Smailagic:2003rp,Smailagic:2003yb,Smailagic:2004yy}. 
The key ingredient of this development
is to define the fields as mean value over coherent states of the
noncommutative plane, such that a star product needs not be introduced.
More recently, it has been shown that the coherent state approach can
be summarized through the introduction of a new multiplication
rule which is known as Voros star-product \cite{Galluccio:2008wk,Galluccio:2009ss}, 
\cite{Banerjee:2009xx,Gangopadhyay:2010zm,Basu:2011kh}.
Anyway, physics turns out be independent from the choice of the type
of product \cite{Hammou:2001cc}.
An alternative view of these modifications  is to consider them
as a redefinition of the Fourier transform of the fields. As a consequence,
the theory is ultraviolet finite and the cutoff is provided by the noncommutative parameter $\theta$. 
It must be clear from this discussion that the existence of a minimal length is determined by the noncommutative
parameter $\theta$. Indeed, since one can incorporate a minimal
length ($\sqrt{\theta}$) in spacetime by assuming nontrivial coordinate
commutation relations, we then have introduced a noncommutative geometry.
Interestingly enough, every point-like structure is smeared out by the
presence of the new quantum of length in the manifold.

On the other hand, in previous studies \cite{Gaete:2004ga,Gaete:2007ax},we have considered
 both axionic electrodynamics and its noncommutative version with a Lorentz invariance violating term, 
in the presence of a nontrivial
constant expectation value for the gauge field strength. This noncommutative
version was discussed to leading order in $\theta$, via the Seiberg-Witten
map. We note that these theories experience mass generation due to the
breaking of rotational invariance induced by the classical background
configuration of the gauge field strength, and in the case of a constant
electric field strength expectation value the static potential  remains
Coulombic for both theories. Nevertheless, this picture drastically changes
in the case of a constant magnetic field strength expectation value.
In effect, for axionic electrodynamics the potential energy is the sum of
a Yukawa and a linear potential, leading to the confinement
of static charges. While for noncommutative axionic electrodynamics the
interaction energy is the sum of a Coulomb and a linear potential.
Nevertheless, the above models are not ultraviolet finite theories.

Inspired by the above observations, one then naturally asks whether there
is a consistent and well-defined axionic electrodynamics in the presence
of a minimal length. It appears that this is indeed so as we show in this
paper. More specifically, the main goal of this work shall be to examine
the effect of the spacetime noncommutativity on a physical observable.
To do this, we will work out the static potential for the theory under
consideration by using the gauge-invariant but path-dependent variables
formalism, which is alternative to the Wilson loop approach \cite{Gaete:2007zn}. 
Our treatment is exact for the noncommutative parameter $\theta$, in other words, 
there is no $\theta$ expansion. Interestingly, our analysis
leads to a well-defined noncommutative interaction energy. In fact, we
obtain an ultraviolet finite static potential which is the sum of a
Yukawa-type and a linear potential, leading to the confinement of static
charges. The key role played by the new quantum of length in triggering a
well-defined interaction energy is our main result.

\section{Finite electrodynamics}

As stated in the introduction, the main focus of this paper
is to reexamine the interaction energy between static point-like
sources for noncommutative axionic electrodynamics. To carry out such
study, we shall compute the expectation value of the energy operator $H$
in the physical state $|\Phi\rangle$, which we will denote by ${\langle
H\rangle}_\Phi$. However, the analysis will be first carried out for a
noncommutative version of electrodynamics to grasp more easily the central
features and will then be extended to noncommutative axionic electrodynamics.

Before starting our analysis it may be useful to recall as
effective Lagrangian look like in the non-commutative coordinates coherent
state approach.  We mean \textit{effective} because they are ordinary
quantum field theory defined over a smooth, flat
\cite{Smailagic:2003rp,Smailagic:2003yb,Smailagic:2004yy}
or curved manifold
\cite{Nicolini:2005de,Nicolini:2005vd,Spallucci:2006zj,Ansoldi:2006vg},
\cite{Spallucci:2008ez,Nicolini:2009gw,Smailagic:2010nv,Spallucci:2011rn},
\cite{Rinaldi:2010zu,Nicolini:2009dr,Rinaldi:2009ba},
with a ``memory'' of the non-classical nature of the coordinates below a
certain length scale. In a series of papers
\cite{Smailagic:2003rp,Smailagic:2003yb,Smailagic:2004yy} it has been shown
how non-commutative geometry effects are recorded in a long-distance
quantum field theory. We can summarize the final results as follows:
\begin{enumerate}
\item point-like particles have physical meaning. Any physical
object has linear size $l\ge \sqrt\theta$\label{uno}.
\item It follows from [\ref{uno}] that there are no point-like sources.
Thus, distributions like $J\propto \delta\left(\, \vec{x}\,\right)$
have to be replaced by \textit{minimal width Gaussian} functions. From
a formal point of view, such a smearing is implemented
through the substitution rule

\begin{equation}
\delta\left(\, \vec{x}\,\right)\longrightarrow
e^{\theta\nabla^2}\delta\left(\, \vec{x}\,\right).
\end{equation}
\label{due}
\item
Smeared sources lead to ultraviolet suppressed (euclidean) Feynman
propagators of the form
\begin{equation}
G\left(\, k^2\,\right)=\frac{ e^{-\theta\, k^2}}{k^2+m^2}.
\label{ncprop}
\end{equation}

This kind of propagators can be obtained from modified kinetic terms.
In the simplest case of a scalar particle the effective Lagrangian
leading to (\ref{ncprop}) reads

\begin{equation}
{\cal L}=\frac{1}{2}\left(\, \partial_\mu \phi\,\right)\, e^{\theta\Delta}
\,\left(\,\partial^\mu\, \phi\,\right)
+\frac{m^2}{2} \phi\, e^{\theta\Delta}\,\phi
\label{ncscalar}
\end{equation}
where, $\Delta  \equiv \partial _\mu  \partial ^\mu$ \label{tre}.
\end{enumerate}
This is a model with derivatives of arbitrary order. In this paper we are
going to investigate \emph{static} potential between test charges, thus
the D'Alembert operator will be replaced by the Laplace operator and only
spatial derivatives will appear. From this point of view, our work is
complementary the  \cite{Moeller:2002vx} where, in a p-adic string model
dilaton dynamics is considered is described by a similar, exponential operator
in time derivatives. Both in  \cite{Moeller:2002vx} and in the present
work the fully covariant case with the exponential of the D'Alembert operator
is not addressed. It is well beyond the purpose of this paper, which is focused
on the static potential, and would deserve an in depth analysis by itself.
Some details on the perturbative treatment of models with higher order derivatives
are given in Appendix A and Appendix B.\\
It is not difficult to guess that the effective Maxwell Lagrangian density
turns out to be
\begin{equation}
\mathcal{L} =
- \frac{1}{4}F_{\mu \nu }\, e^{\theta \Delta } F^{\mu \nu }. \label{NCMaxwell5}
\end{equation}

We now discuss  the interaction energy between static pointlike
sources for this noncommutative electrodynamics, through two different
methods. The first approach is based on the path-integral formalism,
whereas the second one makes use of the gauge-invariant but
path-dependent variables formalism. 

Let us start by writing down the functional generator of the Green's functions, that is,
\begin{equation}
Z\left[ J \right] = \exp \left( { - \frac{i}{2}\int {d^4 xd^4 y}
J^\mu \left( x \right)D_{\mu \nu } \left( {x,y} \right)J^\nu  \left(
y \right)} \right), \label{NCMaxwell5-1}
\end{equation}
where $D_{\mu \nu } \left( {x,y} \right) = \int {\frac{{d^4
k}}{{\left( {2\pi } \right)^4 }}} D_{\mu \nu } \left( k \right)e^{ -
ikx}$ is the propagator in the Feynman gauge. In this case, the
corresponding propagator is given by
\begin{equation}
D_{\mu \nu } \left( k \right) =  - \frac{1}{{k^2 }}\left\{ {e^{\theta k^2 } \eta _{\mu \nu }  + 
\left( {1 - e^{\theta k^2 } } \right)\frac{{k_\mu  k_\nu  }}{{k^2 }}} \right\}.
\label{NCMaxwell5-2}
\end{equation}
By means of expression $Z = e^{iW\left[ J \right]}$, and employing
Eq. (\ref {NCMaxwell5-1}), $W\left[ J \right]$ takes the form
\begin{equation}
W\left[ J \right] =  - \frac{1}{2}\int {\frac{{d^4 k}}{{\left( {2\pi } \right)^4 }}} J_\mu ^ *  
\left( k \right)\left[ { - \frac{{e^{\theta k^2 } }}{{k^2 }}\eta ^{\mu \nu } 
 - \frac{{\left( {1 - e^{\theta k^2 } } \right)}}{{k^2 }}\frac{{k^\mu  k^\nu  }}{{k^2 }}} \right]J_\nu  
\left( k \right).  \label{NCMaxwell5-3}
\end{equation}
Since the current $J^\mu (k)$ is conserved, expression
(\ref{NCMaxwell5-3}) then becomes
\begin{equation}
W\left[ J \right] = \frac{1}{2}\int {\frac{{d^4 k}}{{\left( {2\pi } \right)^4 }}} J_\mu ^ *  
\left( k \right)\left( {\frac{{e^{\theta k^2 } }}{{k^2 }}} \right)J^\mu  \left( k \right).
\label{NCMaxwell5-4}
\end{equation}
Next, for $J_\mu  \left( x \right) = \left[ {Q\delta ^{\left( 3
\right)} \left( {{\bf x} - {\bf x}^{\left( 1 \right)} } \right) + Q^
\prime  \delta ^{\left( 3 \right)} \left( {{\bf x} - {\bf x}^{\left(
2 \right)} } \right)} \right]\delta _\mu ^0$, and using standard
functional techniques \cite{Zee}, we obtain that the interaction
energy of the system is given by
\begin{equation}
V\left( r \right) = QQ^ \prime  \int {\frac{{d^3 k}}{{\left( {2\pi } \right)^3 }}
\frac{{e^{ - \theta {\bf k}^2 } }}{{{\bf k}^2 }}} e^{i{\bf k} \cdot {\bf r}} , \label{NCMaxwell5-5}
\end{equation}
where $ {\bf r} \equiv {\bf x}^{\left( 1 \right)}  - {\bf x}^{\left(
2 \right)}$. 
In order to calculate $V(r)$ we note that the integral over ${\bf k}$ can also
be written as
\begin{eqnarray}
\int {\frac{{d^3 k}}{{\left( {2\pi } \right)^3 }}\frac{{e^{ - \theta {\bf k}^2 } }}{{{\bf k}^2 }}} e^{i{\bf k} 
\cdot {\bf r}}  &=& \int {\frac{{d^3 k}}{{\left( {2\pi } \right)^3 }}} \int\limits_0^\infty 
 {ds} e^{ - \left( {\theta  + s} \right){\bf k}^2 } e^{i{\bf k} \cdot {\bf r}} \nonumber\\
&=&\frac{1}{{4\left( \pi  \right)^{{\raise0.5ex\hbox{$\scriptstyle 3$}
\kern-0.1em/\kern-0.15em
\lower0.25ex\hbox{$\scriptstyle 2$}}} }}\frac{1}{r}\gamma
\left( {{\raise0.5ex\hbox{$\scriptstyle 1$}
\kern-0.1em/\kern-0.15em
\lower0.25ex\hbox{$\scriptstyle 2$}};{\raise0.5ex\hbox{$\scriptstyle {r^2 }$}
\kern-0.1em/\kern-0.15em
\lower0.25ex\hbox{$\scriptstyle {4\theta }$}}} \right), \label{NCMaxwell5-6}
\end{eqnarray}
with $r = |{\bf r}|$. Here $\gamma \left( {{\raise0.5ex\hbox{$\scriptstyle 1$}
\kern-0.1em/\kern-0.15em\lower0.25ex\hbox{$\scriptstyle 2$}};
{\raise0.5ex\hbox{$\scriptstyle {r^2 }$}\kern-0.1em/\kern-0.15em
\lower0.25ex\hbox{$\scriptstyle {4\theta }$}}} \right)$ is the lower
incomplete Gamma function defined by the following integral
representation
\begin{equation}
\gamma \left(\, \frac{a}{b}\ ; x\, \right) \equiv \int_0^x {\frac{{du}}{u}}\,
u^{{a \mathord{\left/
{\vphantom {a b}} \right.
\kern-\nulldelimiterspace} b}}\, e^{ - u}.
\label{NCMaxwell5-7}
\end{equation}
By means of expression (\ref{NCMaxwell5-6}) together with $Q^\prime=-Q$, the interaction energy reduces to 
\begin{equation}
V\left( r \right) = 
 -  \frac{Q^2}{{4\left( \pi  \right)^{{\raise0.5ex\hbox{$\scriptstyle 3$}\kern-0.1em/\kern-0.15em
\lower0.25ex\hbox{$\scriptstyle 2$}}} }}\frac{1}{r}\gamma
\left( {{\raise0.5ex\hbox{$\scriptstyle 1$}
\kern-0.1em/\kern-0.15em
\lower0.25ex\hbox{$\scriptstyle 2$}};{\raise0.5ex\hbox{$\scriptstyle {r^2 }$}
\kern-0.1em/\kern-0.15em
\lower0.25ex\hbox{$\scriptstyle {4\theta }$}}} \right). \label{NCMaxwell5-8}
\end{equation}
From this expression it should be clear that
the interaction energy is regular at the origin, in contrast to the
usual Maxwell theory. In this respect the above result clearly shows
the key role played by the "smeared propagator" in Eq. (\ref{NCMaxwell5-4}) .

Next we compute the interaction energy from the viewpoint of the gauge-invariant but path-dependent variables 
formalism, along the lines of Refs. \cite{Gaete:2004ga,Gaete:2007ax,Gaete:2007zn}. 
Within this framework, we shall compute the expectation value of the energy operator $H$ in the physical 
state $|\Phi\rangle$, which we will denote by ${\langle H \rangle}_\Phi$.  Nevertheless, to obtain the 
corresponding Hamiltonian we must carry out the quantization of the theory. 
At this point, special care has to be exercised since expression (\ref{NCMaxwell5}) contains higher time 
derivatives.
However, as was mentioned before, this paper is aimed at studying
the static potential of the above theory, so that $\Delta$ can be
replaced by $ - \nabla ^2$. At the moment for notational convenience
we will maintain $\Delta$, but it should be borne in mind that this
paper essentially deals with the static case. In addition, it is interesting
to note that if we expand expression (\ref{NCMaxwell5}) up to first
order in $\theta$, we  obtain the Lagrangian density of the Abelian
Lee-Wick model \cite{ Accioly:2011zz,Accioly:2010js}. We shall come back to these points
in the appendix A. 

We now turn our attention to the calculation
of the interaction energy. In order to obtain the corresponding Hamiltonian,
the canonical quantization of this theory from the Hamiltonian point of
view is straightforward. The canonical momenta are found to be
 $ \Pi ^\mu   =  - e^{\theta \Delta } F^{0\mu }$, and one
immediately identifies the usual primary constraint $\Pi ^0  = 0$
and $\Pi ^i  = e^{\theta \Delta } F^{i0}$. The canonical Hamiltonian
is now obtained in the usual way by a Legendre transform, that is,

\begin{equation}
H_C  = \int {d^3 x} \left\{ { - A_0 \partial _i \Pi ^i  -
\frac{1}{2}\Pi _i e^{- \theta \Delta } \Pi ^i  + \frac{1}{4}
F_{ij} e^{\theta \Delta } F^{ij} } \right\}.
\label{NCMaxwell10}
\end{equation}

Time conservation of the primary constraint, $\Pi _0  = 0$, leads
to the usual Gauss constraint
$\Gamma_1 \left( x \right) \equiv \partial _i \Pi ^i=0$.
The extended Hamiltonian that generates translations in time then
reads $H = H_C + \int {d^2}x\left( {c_0 \left( x \right)\Pi _0
\left( x \right) + c_1 \left(x\right)\Gamma _1 \left( x \right)}
\right)$, where $c_0 \left(x\right)$ and $c_1 \left( x \right)$ are
the Lagrange multipliers. Since $ \Pi^0 = 0$ for all time and
$\dot{A}_0 \left( x \right)= \left[ {A_0 \left( x \right),H} \right] =
c_0 \left( x \right)$, which is completely arbitrary, we discard
$A^0$ and $\Pi^0$ because they adding nothing to the description of
the system. Thus the extended Hamiltonian is now given as

\begin{equation}
H = \int {d^3 x} \left\{ {c(x)\partial _i \Pi ^i  - \frac{1}{2}
\Pi _i e^{- \theta \Delta } \Pi ^i  + \frac{1}{4}F_{ij}
e^{\theta \Delta } F^{ij} } \right\}
, \label{NCMaxwell15}
\end{equation}
where $c(x) = c_1 (x) - A_0 (x)$ is an arbitrary parameter reflecting
the gauge invariance of the theory. As is well known, to avoid this
trouble we must fix the gauge. A particularly convenient choice is found
to be
\begin{equation}
\Gamma _2 \left( x \right) \equiv \int\limits_{C_{\xi x} } {dz^\nu }
A_\nu \left( z \right) \equiv \int\limits_0^1 {d\lambda x^i } A_i
\left( {\lambda x} \right) = 0,     \label{NCMaxwell20}
\end{equation}
where  $\lambda$ $(0\leq \lambda\leq1)$ is the parameter describing
the space-like straight path $ x^i = \xi ^i  + \lambda \left( {x -
\xi } \right)^i $, and $ \xi $ is a fixed point (reference point).
There is no essential loss of generality if we restrict our
considerations to $ \xi ^i=0 $. The choice (\ref{NCMaxwell20}) leads to
the Poincar\'e gauge \cite{Gaete:1997eg}. As was explained in \cite{Gaete:1997eg}, 
we can now write down the only non-vanishing Dirac bracket for the canonical variables.
This is a fairly long calculation which will not repeat here.
\begin{eqnarray}
\left\{ {A_i \left( x \right),\Pi ^j \left( y \right)} \right\}^ *
=\delta{ _i^j} \delta ^{\left( 3 \right)} \left( {x - y} \right) 
-\partial _i^x \int\limits_0^1 {d\lambda x^j } \delta ^{\left( 3
\right)} \left( {\lambda x - y} \right). \label{NCMaxwell25}
\end{eqnarray}

We now proceed with the calculation of the interaction energy between
point-like sources for the model under consideration. As we have noted
before, we will calculate the expectation value of the energy operator
$H$ in the physical state $|\Phi\rangle$. At this point, we also recall
that the physical state $|\Phi\rangle$ can be written as
\begin{eqnarray}
\left| \Phi  \right\rangle  \equiv \left| {\overline \Psi  \left(
\bf y \right)\Psi \left( {\bf y}\prime \right)} \right\rangle 
= \overline \psi \left( \bf y \right)\exp \left(
{iq\int\limits_{{\bf y}\prime}^{\bf y} {dz^i } A_i \left( z \right)}
\right)\psi \left({\bf y}\prime \right)\left| 0 \right\rangle,
\label{NCMaxwell30}
\end{eqnarray}
where the line integral is along a spacelike path on a fixed time
slice, $q$ is the fermionic charge, and $\left| 0 \right\rangle$ is
the physical vacuum state. Note that the charged matter field together
with the electromagnetic cloud (dressing) which surrounds it, is given
by $\Psi \left( {\bf y} \right) = \exp \left( { - iq\int_{C_{{\bf \xi}
{\bf y}} } {dz^\mu A_\mu  (z)} } \right)\psi ({\bf y})$. With the help
of our path choice, this physical fermion then becomes $\Psi \left(
{\bf y} \right) = \exp \left( { - iq\int_{\bf 0}^{\bf y} {dz^i  }
A_{i} (z)} \right)\psi ({\bf y})$. In other words, each of the
states ($\left| \Phi  \right\rangle$) represents a
fermion-antifermion pair surrounded by a cloud of gauge fields to
maintain gauge invariance.

Next, by taking into account the above Hamiltonian structure, we
observe that

\begin{eqnarray}
\Pi _i \left(\, x\, \right)\left| {\overline \Psi  \left(\, \mathbf{ y}\,
\right)\Psi \left( \,{{\bf y}^ \prime  } \,\right)} \right\rangle  =
\overline \Psi  \left( \bf y \right)\Psi \left( {{\bf y}^ \prime }
\right)\Pi _i \left( x \right)\left| 0 \right\rangle 
+q\int_ {\bf y}^{{\bf y}^ \prime  } {dz_i \delta ^{\left( 3
\right)} \left( {\bf z - \bf x} \right)} \left| \Phi \right\rangle. \nonumber\\
\label{NCMaxwell31}
\end{eqnarray}

Having made this observation and since the fermions are taken to be
infinitely massive (static) we can substitute $\Delta$ by
$-\nabla^{2}$ in Eq. (\ref{NCMaxwell15}). Therefore, the expectation
value $\left\langle H \right\rangle _\Phi$ becomes
\begin{equation}
\left\langle H \right\rangle _\Phi   = \left\langle H \right\rangle
_0 + \left\langle H \right\rangle _\Phi ^{\left( 1 \right)},
\label{NCMaxwell35}
\end{equation}    
where $\left\langle H \right\rangle _0  = \left\langle 0
\right|H\left| 0 \right\rangle$. The $\left\langle H \right\rangle
_\Phi ^{\left( 1 \right)}$ term is given by

\begin{eqnarray}
\left\langle H \right\rangle _\Phi ^{\left( 1 \right)}  =
- \frac{{q^2 }}{2}
\int {d^3 x} \int_{\bf y}^{{\bf y}^ \prime} {dz_i^ \prime}
\delta ^{\left( 3 \right)}
\left( {{\bf x} - {\bf z}^ \prime} \right)e^{ \theta \nabla _x^2 } 
\int_{\bf y}^{{\bf y}^ \prime}{dz_i^ \prime} 
\delta ^{\left( 3 \right)}\left( {{\bf x} - {\bf z}} \right),
\label{NCMaxwell40a}
\end{eqnarray} \\
which can also be expressed solely in terms of the new Green function

\begin{equation}
\left\langle H \right\rangle _\Phi ^{\left( 1 \right)}  =
- \frac{{q^2 }}
{2}\int_{\bf y}^{{\bf y}^ \prime} {dz^{\prime i} }
\int_{\bf y}^{{\bf y}^ \prime} {dz^i } \nabla _z^2
\widetilde G \left( {{\bf z},{\bf z}^ \prime} \right).
\label{NCMaxwell40b}
\end{equation}
In this case, $\widetilde G$ is the new Green function

\begin{equation}
\widetilde G \left({{\bf z},{\bf z}^ \prime} \right) =
\frac{1}{{4\left( \pi  \right)^{{\raise0.5ex\hbox{$\scriptstyle 3$}
\kern-0.1em/\kern-0.15em
\lower0.25ex\hbox{$\scriptstyle 2$}}} }}\frac{1}{r}\gamma
\left( {{\raise0.5ex\hbox{$\scriptstyle 1$}
\kern-0.1em/\kern-0.15em
\lower0.25ex\hbox{$\scriptstyle 2$}};{\raise0.5ex\hbox{$\scriptstyle {r^2 }$}
\kern-0.1em/\kern-0.15em
\lower0.25ex\hbox{$\scriptstyle {4\theta }$}}} \right),
\label{NCMaxwell40c}
\end{equation}
where $r \equiv |{\bf z} -{\bf z}^{\prime}|$.

Employing  Eq.(\ref{NCMaxwell40b}) and remembering that the integrals
over ${z^i}$ and ${z^{\prime i}}$ are zero except on the contour of
integration, the potential for two opposite charges, located at
${\bf y}$ and ${\bf y^{\prime}}$, reduces to the Coulomb-type potential.
In other words,

\begin{equation}
V(L) =  - \frac{{q^2 }}{{4\left( \pi  \right)^{{\raise0.5ex\hbox{$\scriptstyle 3$}
\kern-0.1em/\kern-0.15em
\lower0.25ex\hbox{$\scriptstyle 2$}}} }}\frac{1}{L}\,
\gamma
\left(\, 1/2\ ;  L^2/4\theta\, \right)\ ,
\label{NCMaxwell45}
\end{equation}
with $|{\bf y} -{\bf y}^{\prime}|\equiv L$. One immediately sees that both
approaches, despite being completely different, lead to the same result
which seems indicate that they are equivalent term by term. Incidentally,
it is of interest to notice that the above result comes from the constraints structure of the theory under 
consideration. Furthermore, in contrast to our
previous analysis \cite{Gaete:2004ga,Gaete:2007ax} via the Seiberg-Witten map, unexpected features are found. 
Interestingly, it should be noted that the
above calculation of $V$ involves no $\theta$ expansion at all. Note also
that setting the noncommutative parameter to zero reproduces the standard
expression for Maxwell theory. In fact, it is observed that the
introduction of the noncommutative space induces a finite Coulombic
potential for $L \rightarrow 0$ (See Fig. $1$). This then implies that
the self-energy and the electromagnetic mass of a point-like particle are
finite in this version noncommutative of electrodynamics. Thus, the present
theory is extremely simple but still rich in content and its static potential
is remarkably similar to the one found in the Abelian Lee-Wick model.
In addition, it is worthwhile noticing that the potential (\ref{NCMaxwell45})
is spherically symmetric, although the external fields break the isotropy of
the problem in a manifest way. Also, we stress that the choice of the gauge
is in this approach really arbitrary. Being the formalism completely gauge
invariant, we would obtain exactly the same result in any gauge.

\begin{figure}[h]
\begin{center}
\includegraphics[scale=1.5]{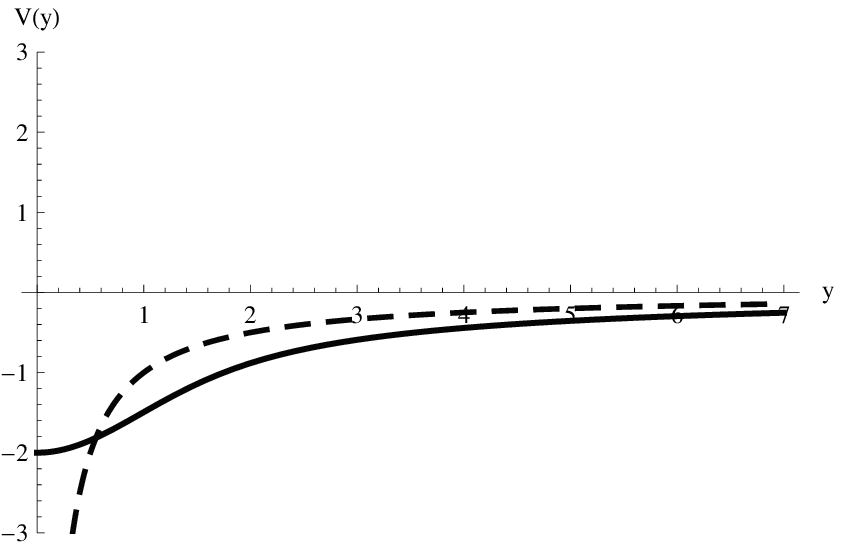}
\end{center}
\caption{\small
Shape of the potential, Eq.(\ref{NCMaxwell45}). Note that
$V\left( L \right) \equiv \frac{{q^2 }}{{8\pi }}\frac{1}
{{\sqrt {\pi \theta } }}V(y)$, with
$y \equiv {\raise0.5ex\hbox{$\scriptstyle L$}
\kern-0.1em/\kern-0.15em
\lower0.25ex\hbox{$\scriptstyle {2\sqrt \theta  }$}}$.
The dashed line represents the Coulomb potential.}
\label{fig1}
\end{figure}

We shall next discuss two alternative derivations of the result
(\ref{NCMaxwell45}). Instead of working with the Lagrangian density
(\ref{NCMaxwell5}), we might as well formulate the discussion in terms
of smeared sources, as was first demonstrated in
\cite{Smailagic:2003rp,Smailagic:2003yb,Smailagic:2004yy}. It also
permits to check the internal consistency of our methodology.

As a first step, we shall begin by considering the usual Lagrangian
density for the Maxwell field, namely,
\begin{equation}
\mathcal{L} =  - \frac{1}{4}F_{\mu \nu } F^{\mu \nu }. \label{NCMaxwell50}
\end{equation}
Following the same steps employed for obtaining (\ref{NCMaxwell35}), the
above $\left\langle H \right\rangle _\Phi ^{\left( 1 \right)}$ term is
expressed as
\begin{eqnarray}
\left\langle H \right\rangle _\Phi ^{\left( 1 \right)}  =
- \frac{{q^2 }}{2}
\int {d^3 x} \int_{\bf y}^{{\bf y}^ \prime} {dz_i^ \prime}
\delta ^{\left( 3 \right)}
\left( {{\bf x} - {\bf z}^ \prime} \right) 
\int_{\bf y}^{{\bf y}^ \prime}{dz_i^ \prime} 
\delta ^{\left( 3 \right)} \left( {{\bf x} - {\bf z}} \right).
\label{NCMaxwell55}
\end{eqnarray}
We now consider the formulation of this theory in the presence of a
minimal length. To do this, the source $\delta ^{\left( 3 \right)}
\left( {x - y} \right)$ is replaced by the smeared source
$e^{\frac{\theta }
{2}\nabla ^2 } \delta ^{\left( 3 \right)} \left( {x - y} \right)$.
Hence expression (\ref{NCMaxwell31}) reduces to
\begin{eqnarray}
\Pi _i \left( x \right)\left| \Phi  \right\rangle  \equiv \overline \Psi
\left( {\bf y} \right)\Psi \left( {{\bf y}^ \prime} \right)\Pi _i
\left| 0 \right\rangle  
+ q\int_{\bf y}^{{\bf y}^ \prime} {dz_i }
e^{\frac{\theta }{2}\nabla ^2 } \delta ^{\left( 3 \right)}
\left( {{\bf z} - {\bf x}} \right) \left( x \right)\left| \Phi  \right\rangle.\label{NCMaxwell60}
\end{eqnarray}
Thus, by employing relation (\ref{NCMaxwell60}) into Eq.(\ref{NCMaxwell55}),
we immediately recover the result (\ref{NCMaxwell45}). 
However, the key point in our analysis is the presence of a smeared source. 
This smearing factor is an intrinsic characteristic of spacetime and not an arbitrary number related to 
particular coordinates. In other terms, the new proposal provides a rich framework in order to include quantum 
effects of spacetime itself.

As a second derivation of our previous result, it may be recalled that
\cite{Gaete:1997eg}:
\begin{equation}
V \equiv q\left( {\mathcal{A}_0 \left( {\bf y} \right) - \mathcal{A}_0
\left( {\bf y \prime} \right)} \right), \label{NCMaxwell70}
\end{equation}
where the physical scalar potential is given by
\begin{equation}
\mathcal{A}_0 \left( {x^0 ,{\bf x}} \right) = \int_0^1 {d\lambda } x^i
E_i \left( {\lambda {\bf x}} \right), \label{NCMaxwell75}
\end{equation}
with $i=1,2,3$. This follows from the vector gauge-invariant field
expression
\begin{equation}
\mathcal{A}_\mu  \left( x \right) \equiv A_\mu  \left( x \right) +
\partial _\mu  \left( { - \int_\xi ^x {dz^\mu  } A_\mu  \left( z
\right)} \right), \label{NCMaxwell80}
\end{equation}
where, as in Eq.(\ref{NCMaxwell30}), the line integral is along a
space-like path from the point $\xi$ to $x$, on a fixed slice time.
The gauge-invariant variables (\ref{NCMaxwell80}) commute with the
sole first constraint (Gauss' law), corroborating that these fields
are physical variables \cite{Dirac}. In passing we note that Gauss' law
for the present theory reads $\partial _i \Pi ^i  = J^0$, where we have
included the external current $J^0$ to represent the presence of two
opposite charges. For, $J^0 \left( {\bf x} \right) = qe^{\theta \nabla ^2 }
\delta ^{\left( 3 \right)} \left( {\bf x} \right)$, we then have that the
electric field may be written as
\begin{equation}
E^i  = q\,\partial ^i \widetilde G \left( x \right).\label{NCMaxwell85}
\end{equation}
Finally, replacing this result in (\ref{NCMaxwell75}) and using
(\ref{NCMaxwell70}), the potential for a pair of point-like opposite charges
q, located at ${\bf 0}$ and ${\bf L}$, takes the form
\begin{equation}
V(L) =  - \frac{{q^2 }}{{4\left( \pi  \right)^{{\raise0.5ex\hbox{$\scriptstyle 3$}
\kern-0.1em/\kern-0.15em
\lower0.25ex\hbox{$\scriptstyle 2$}}} }}\frac{1}{L}\gamma
\left(\, 1/2\ ; L/4\theta\,  \right)\ ,
\label{NCMaxwell90}
\end{equation}
where $\left| \mathbf{L} \right| \equiv L$. It must be clear from this
discussion that a correct identification of physical degrees of freedom
is a key feature for understanding the physics hidden in gauge theories.
According to this viewpoint, once that identification is made, the
computation of the potential is carried out by means of Gauss's law.

\section{Finite axionic electrodynamics}

We now extend our analysis for considering axionic electrodynamics
\cite{Gaete:2004ga}, which is the main thrust of this work. The gauge theory
we are considering is defined by
\begin{equation}
\mathcal{L} =  - \frac{1}{4}F_{\mu \nu } F^{\mu \nu }  + \frac{1}{2}
\left( {\partial _\mu  \varphi } \right)^2  - \frac{1}{2}m^2
\varphi ^2  + \frac{\lambda }{4}\varphi \widetilde F^{\mu \nu }
F_{\mu \nu }.
\label{NCMaxwell95}
\end{equation}
\label{NCMaxwell900}

However, in order to put our discussion into context it is useful to
summarize the relevant aspects of the analysis described previously
\cite{Gaete:2004ga}. Thus, our first undertaking is to carry out the
integration over the $\varphi$-field. We also recall that in
\cite{Gaete:2004ga} we have considered static scalar fields,
as a consequence we may replace  $\Delta\varphi = - \nabla ^2\varphi$,
with $\triangle \equiv \partial^\mu\partial_\mu$. Once this is done,
we arrive at the following effective theory
 \begin{equation}
\mathcal{L} =  - \frac{1}{4}F_{\mu \nu } F^{\mu \nu }  - \frac{{\lambda ^2 }}
{{32}}\left( { \widetilde F_{\mu \nu } F^{\mu \nu } } \right)\frac{1}
{{\nabla ^2  - m^2 }}\left( { \widetilde F_{\alpha \beta }
F^{\alpha \beta } } \right).
\label{NCMaxwell955}
\end{equation}
Next, after splitting $F_{\mu \nu }$ in the sum of a classical background,
$\langle F_{\mu \nu }\rangle$, and a small fluctuation, $f_{\mu \nu }$,
the corresponding Lagrangian density up to quadratic
terms in the fluctuations, becomes
\begin{equation}
\mathcal{L} =  - \frac{1}{4}f_{\mu \nu } f^{\mu \nu }  - \frac{{\lambda ^2 }}
{{32}}v^{\mu \nu } f_{\mu \nu } \frac{1}{{\nabla ^2  - m^2 }}v^{\lambda \rho }
f_{\lambda \rho },
\label{NCMaxwell100}
\end{equation}
where $f_{\mu\nu} = \partial_\mu a_\nu - \partial_\nu a_\mu$, $a_{\mu}$
stands for the fluctuation, and $\varepsilon ^{\mu \nu \alpha \beta }
\left\langle {F_{\alpha \beta } } \right\rangle  \equiv v^{\mu \nu }$
and $\varepsilon ^{\rho\lambda\gamma\delta } \left\langle {F_{\gamma\delta } }
\right\rangle \equiv v^{\rho\lambda }$.

Having made this observation, we now turn our attention to the
calculation of the interaction energy in the
$v^{0i}  \ne 0$ and $v^{ij} = 0$ case (referred to as the electric one
in what follows). In such a case, the Lagrangian (\ref{NCMaxwell100})
reads
\begin{equation}
\mathcal{L} =  - \frac{1}{4}f_{\mu \nu } f^{\mu \nu }  - \frac{{\lambda ^2 }}
{{32}}v^{0i} f_{0i} \frac{1}{{\nabla ^2  - m^2 }}v^{0k} f_{0k}.
\label{NCMaxwell105}
\end{equation}

It is now again straightforward to apply the formalism discussed in the
preceding section. For this purpose, we start by observing that the
canonical Hamiltonian can be worked as usual and is given by
\begin{equation}
H_C  = \int {d^3 x} \left\{ {\Pi _i \partial ^i a^0  + \frac{1}{2}\Pi ^i
\frac{{\left( {\nabla ^2  - m^2 } \right)}}{{\left( {\nabla ^2  - M^2 } \right)}}
\Pi ^i  + \frac{1}{2}{\bf B}^2 } \right\},
\label{NCMaxwell110}
\end{equation}
where $\bf B$ is the magnetic field, and $M^2  = m^2  + \frac{{\lambda ^2 }}
{16}{\bf v}^2$.

As we pointed before, by employing relation (\ref{NCMaxwell60}), the expectation value reads,
\begin{equation}
\left\langle H \right\rangle _\Phi   = \left\langle H \right\rangle _0
+ \left\langle H \right\rangle _\Phi ^{\left( 1 \right)}  + \left\langle H
\right\rangle _\Phi ^{\left( 2 \right)},
\label{NCMaxwell115}
\end{equation}
with  $\left\langle H \right\rangle _0  = \left\langle 0 \right|H\left| 0
 \right\rangle$,
while the terms $\left\langle H \right\rangle _\Phi ^{\left( 1 \right)}$ and
$\left\langle H \right\rangle _\Phi ^{\left( 2 \right)}$ are given by

\begin{eqnarray}
\left\langle H \right\rangle _\Phi ^{\left( 1 \right)}  &=& - \frac{{q^2 }}
{2}\int {d^3 x} \int_{\bf y}^{{\bf y}^ \prime  } {dz_i^ \prime  } e^{
\frac{\theta }{2}\nabla _{z^ \prime  }^2 } \delta ^{\left( 3 \right)}
\left( {{\bf x} - {\bf z}^ \prime  } \right) \nonumber\\
&\times&\left( {1 - \frac{{M^2 }}{{\nabla ^2 }}} \right)_x^{ - 1} 
\int_{\bf y}^{{\bf y}^ \prime  } {dz^i } e^{  \frac{\theta }{2}
\nabla _z^2 } \delta ^{\left( 3 \right)} \left( {{\bf x} - {\bf z}} \right), 
\nonumber\\
\label{NCMaxwell120a}
\end{eqnarray}
and
\begin{eqnarray}
\left\langle H \right\rangle _\Phi ^{\left( 2 \right)}  &=& \frac{{m^2 q^2 }}
{2}\int {d^3 x} \int_{\bf y}^{{\bf y}^ \prime  } {dz_i^ \prime  } e^{
\frac{\theta }{2}\nabla _{z^ \prime }^2 } \delta ^{\left( 3 \right)}
\left( {{\bf x} - {\bf z}^ \prime  } \right) \nonumber\\
&\times&\left( {\frac{1}{{\nabla ^2  - M^2 }}}
\right)_x \int_{\bf y}^{{\bf y}^ \prime  } {dz^i } e^{  \frac{\theta }{2}
\nabla _z^2 } \delta ^{\left( 3 \right)} \left( {{\bf x} - {\bf z}} \right).
\nonumber\\
\label{NCMaxwell120b}
\end{eqnarray}
Following our earlier discussion, these expressions become

\begin{equation}
\left\langle H \right\rangle _\Phi ^{\left( 1 \right)}  =  - \frac{{q^2 }}
{2}\int_{\bf y}^{{\bf y}^ \prime  } {dz^{ \prime i} } \int_{\bf y}^{{\bf y}^ \prime  }
{dz^i } \nabla _z^2 \widetilde G\left( {{\bf z},{\bf z}^ \prime  } \right),
\label{NCMaxwell125}
\end{equation}
and
\begin{equation}
\left\langle H \right\rangle _\Phi ^{\left( 2 \right)}  = \frac{{q^2 m^2 }}
{2}\int_{\bf y}^{{\bf y}^ \prime  } {dz^{ \prime i} } \int_{\bf y}^{{\bf y}^ \prime }
{dz^i } \widetilde G\left( {{\bf z},{\bf z}^ \prime  } \right).
\label{NCMaxwell130} 
\end{equation} 

Accordingly, the new Green function takes the form\\
\begin{eqnarray}
\widetilde{G} =
\frac{e^{M^2 \theta } }{4\pi^{3/2}}
\sqrt{\frac{2M}{r}}\, \Biggl[\, K_{1/2}
\left(\, r\,\right)
-\frac{1}{2}\int_{ r/(2M\theta) }^\infty  dy\, y^{-1/2}
\,e^{ - \frac{Mr}{2}\left(\, y + 1/y\, \right)} \,
\Biggr] \ .
\label{NCMaxwell135}
\end{eqnarray} \\

By using $\sqrt {\frac{\pi }{{2x}}} K_{{\raise0.5ex\hbox{$\scriptstyle 1$}
\kern-0.1em/\kern-0.15em \lower0.25ex\hbox{$\scriptstyle 2$}}}
\left( x \right) = \frac{\pi }{{2x}}e^{ - x}$, it follows that this new Green
function can be rewritten as

\begin{eqnarray}
\widetilde G = \frac{{e^{M^2 \theta } }}{{4\pi }}\frac{1}{r} \Biggl[\, e^{ - Mr} 
- \frac{1}{{\sqrt \pi  }}\int_{r^2/4\theta }^\infty  du\, u^{-1/2}
e^{ - u - (M^2 r^2/ 4u)}\,\Biggr]\ ,
\label{NCMaxwell140}
\end{eqnarray}
which is finite in the limit $r\rightarrow0$, that is,
$\widetilde G\left(\, 0\,\right) =
- \frac{M}{{4\pi }}e^{M^2 \theta }$. \\

Finally, following our earlier procedure, the potential for a pair of
point-like opposite charges q located at ${\bf 0}$ and ${\bf L}$ takes
the form

\begin{eqnarray}
V(L) &=&  - \frac{q^2 }{4\pi }\frac{e^{M^2 \theta } }{L} \Biggl[\, e^{ - ML}
- \frac{1}{\sqrt \pi  }\,\int_{L^2/4\theta}^\infty 
dy\, y^{-1/2}e^{ - y - (M^2 L^2 /4y)}\, \Biggr] \nonumber\\ 
&+& \frac{{q^2 }}{{8\pi }}m^2\, e^{M^2 \theta }\, E_1\, \left( {M^2 \theta }
\right)\, L\ , \label{NCMaxwell1155}
\end{eqnarray}
where $E_1 (M^2 \theta )$ is the exponential integral. Once again, this
result explicitly shows the effect of including a smeared source in the
form of an ultraviolet finite static potential which is the sum of a Yukawa
and a linear potential, leading to the confinement of static charges
(See Fig 2). Another crucial feature of this result is that the entire
effect of noncommutativity is properly captured in the string tension.
This improves the situation as compared to our previous studies
\cite{Gaete:2004ga,Gaete:2007ax}, where an ultraviolet cutoff has been
introduced.
\begin{figure}[h]
\begin{center}
\includegraphics[scale=1.5]{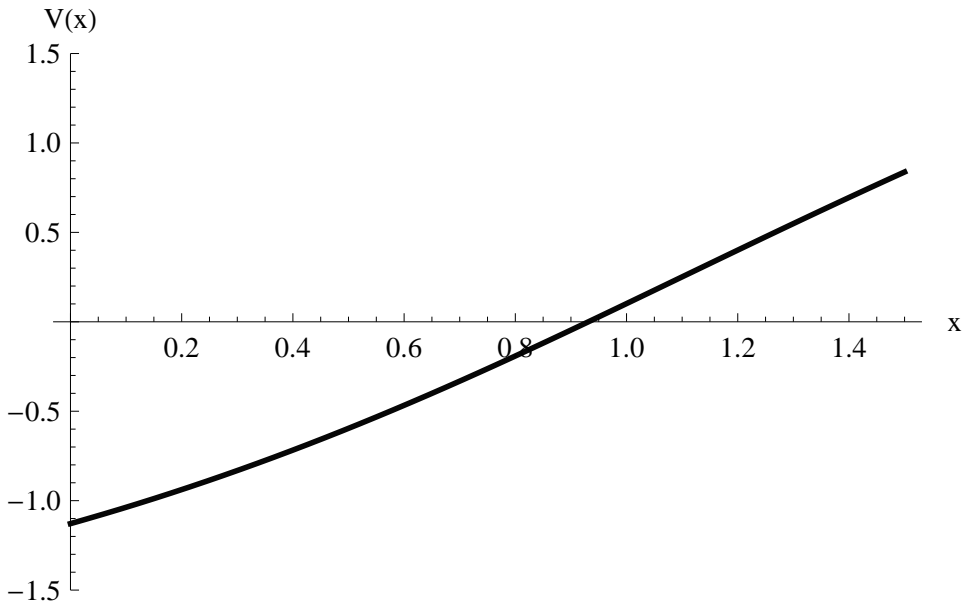}
\end{center}
\caption{\small
Shape of the potential, Eq.(\ref{NCMaxwell1155})}
\label{fig2}
\end{figure}

Note that in figure (\ref{fig2}) we defined
$V(L) = \frac{q^2 e^{M^2 \theta } M}{4\pi }\, V(x)$.
The plot represents the potential energy for the case
$4M^2\theta=0.2 $ and $\frac{m^2}{2M^2}E_1 \left( M^2 \theta \right) =2$.

\section{Final remarks}

In summary, within the gauge-invariant but path-dependent variables formalism,
we have considered the confinement versus screening issue for axionic
electrodynamics in the presence of a minimal length. Once again, a correct
identification of physical degrees of freedom has been a key tool for
understanding the physics hidden in gauge theories.
Interestingly, we have obtained an ultraviolet finite static potential which
is the sum of a Yukawa-type and a linear potential, leading to the confinement
of static charges. This may be contrasted with our previous studies
\cite{Gaete:2004ga,Gaete:2007ax}, where a cutoff has been introduced in order to avoid ultraviolet divergences. 
As already expressed, this calculation involves no $\theta$ expansion at all. The above analysis reveals the key 
role played by the new quantum of length in our analysis. Finally, it seems a challenging work to extend to the
 non-Abelian case as well as to three-dimensional gauge theories the above analysis. We expect to report on 
progress along these lines soon.

\section{Appendix A: Higher time derivatives and Lee-Wick model.}

In this appendix we wish to further elaborate on our previous observation after expression (\ref{NCMaxwell5}), 
that is, by dealing with static configurations we may use the standard Legendre transformation to construct 
the Hamiltonian.

The initial point of our analysis is the Lagrangian density (\ref{NCMaxwell5}),
\begin{equation}
{\cal L} =  - \frac{1}{4}F_{\mu \nu } e^{\theta \Delta } F^{\mu \nu } - J_\mu A^\mu, \label{A1}
\end{equation}
where $J_\mu$ is an external source. In order to illustrate the role played
by higher time derivatives, for simplicity, we consider expression (\ref{A1}) to
lowest order in $\theta$. In this case,
\begin{equation}
{\cal L} =  - \frac{1}{4}F_{\mu \nu } \left( {1 + \theta \Delta } \right)
F^{\mu \nu } - J_\mu A^\mu,\label{A2}
\end{equation}
which is similar to Lee-Wick electrodynamics \cite{Lee-Wick}. However, expression
(\ref{A2}) can also be written as
\begin{equation}
{\cal L} =  - \frac{1}{4}F_{\mu \nu } F^{\mu \nu }  + \frac{\theta }{2}
\partial _\lambda  F^{\lambda \alpha } \partial ^\rho  F_{\rho \alpha } - J_\mu A^\mu, \label{A3}
\end{equation}
which is known as Podolsky's electrodynamics \cite{Podolsky}. In passing we note that
this Lagrangian is the simplest system with second time derivatives.

One immediately sees that the Euler-Lagrange equations are
\begin{equation}
\left( {1 + \theta \Delta } \right)\partial _\mu  F^{\mu \lambda }  =
J^\lambda,\label{A4}
\end{equation}
where $\Delta  \equiv \partial _\mu  \partial ^\mu$.
Expressed in terms of electric and magnetic fields, $E^i  = F^{0i}$ and $ B^i  
= \frac{1}{2}\varepsilon ^{ijk} F_{jk}$, the equations of motion take the form
\begin{equation}
\left( {1 + \theta \Delta } \right)\nabla\cdot{\bf E} = J^0, \label{A5}
\end{equation}
and
\begin{equation}
\left( {1 + \theta \Delta } \right)\left( {\dot {\bf E} - \nabla  \times
{\bf B}} \right) = {\bf J}. \label{A6}
\end{equation}

Interestingly, it is observed that in the electrostatic case ($\dot{\bf E}=0$ and
${\bf B}=0$), and ${\bf J}=0$, these equations reduce to
\begin{equation}
\left( {\nabla ^2  - {\raise0.7ex\hbox{$1$} \!\mathord{\left/
{\vphantom {1 {\theta ^2 }}}\right.\kern-\nulldelimiterspace}
\!\lower0.7ex\hbox{${\theta  }$}}} \right)\left( {\nabla \cdot{\bf E}} \right) = J^0. \label{A7}
\end{equation}

For  $J^0 \left( {t,{\bf x}} \right) = q\delta ^{\left( 3 \right)} \left( {\bf x}
\right)$, the electric field is given by
\begin{equation}
E^i  = q\partial ^i \left( {G\left( {\bf x} \right) + \tilde G\left( {\bf x}
\right)} \right), \label{A8}
\end{equation}
where $G\left( x \right) = \frac{1}{{4\pi }}\frac{1}{{|x|}}$ and
$\tilde G\left( x \right) = \frac{{e^{ - \frac{{|x|}}{{\sqrt \theta}}}}}{{4\pi |x|}}$
are the Green functions in three space dimensions. Next, replacing this result in
Eq. (\ref{NCMaxwell75}), the potential for a pair of point-like opposite charges q
located at ${\bf 0}$ and ${\bf L}$ becomes
\begin{equation}
V =  - \frac{{q^2 }}{{4\pi }}\frac{1}{L}\left( {1 - e^{ - \frac{L}{{\sqrt \theta}}}} \right), \label{A9}
\end{equation}
with $L \equiv |{\bf L}|$. Incidentally, the above static potential is identical to
the one encountered in Ref. \cite{Accioly:2010js} in the Hamiltonian approach, by
using the standard Legendre transformation. But we do not think that the agreement
is an accidental coincidence.\\

In order to understand more precisely this agreement, we will  reexamine our previous Hamiltonian analysis. 
In this case, it is well known that the Hamiltonian approach to
higher derivatives theories was first developed by Ostrogradsky \cite{Ostrogradsky} for non-singular systems, 
and his method consists in defining one more pair of canonical variables and so doubling the dimension of the 
phase space. For singular higher
derivatives systems (our case) one can generalize Dirac's theory for constrained systems
to include the Ostrogradsky approach. Accordingly, in Lagrangian (\ref{A3}) the
velocities have to be taken as independent canonical variables. Hence, the phase-space coordinate 
for the theory under consideration is given by
\begin{equation}
\left( {A_\mu  ,\Pi ^\nu  } \right) \oplus \left( {\dot A_\mu  ,\Pi
^{\left( 1 \right)\nu } } \right), \label{A10}\\
\end{equation}
where $\Pi ^{\left( 1 \right)\nu }$ is the canonical momentum conjugate to
$ \dot A_\mu $. This implies that the canonical Hamiltonian $H_{C}$ takes the form
\begin{equation}
H_C  = \int {d^3 x\left( {p_\mu  \dot A^\mu   + \Pi _\mu ^{\left( 1 \right)}
\ddot A^\mu   - {\cal L}} \right)} . \label{A11}
\end{equation}
According to usual procedure, the momenta are defined as \cite{GalPim}:
\begin{equation}
\Pi _\mu ^{\left( 1 \right)}  \equiv \frac{{\partial {\cal L}}}{{\partial
\left( {\ddot A^\mu  } \right)}}, \label{A12}
\end{equation}
and
\begin{equation}
p_\mu   \equiv \frac{{\partial {\cal L}}}{{\partial \dot A^\mu  }} - 2\partial _k
\left[ {\frac{{\partial {\cal L}}}{{\partial \left( {\partial _0 \partial _k
A^\mu  } \right)}}} \right] - \partial _0 \left( {\frac{{\partial {\cal L}}}
{{\partial \ddot A^\mu  }}} \right). \label{A13}
\end{equation}

Using these definitions, we obtain the following expressions for the momenta:
\begin{equation}
\Pi _\mu ^{\left( 1 \right)}  = \theta \left[ {\partial _\lambda  F^{0\lambda }
\delta _\mu ^0  + \partial ^\lambda  F_{\lambda \mu } } \right], \label{A14}
\end{equation}
and
\begin{equation}
p_\mu   =  - F_{0\mu }  - \theta \left[ {2\partial _k \partial _\lambda
F^{0\lambda } \delta _\mu ^k  - \partial _0 \partial ^\lambda  F_{\mu \lambda}}
\right]. \label{A15}
\end{equation}
Hence, we find
\begin{equation}
{\bf \Pi} ^{\left( 1 \right)}  = \theta \left[ {\dot {\bf E} - \left( {\nabla \times
{\bf B}} \right)} \right], \label{A16}
\end{equation}
and
\begin{equation}
{\bf p} =  - {\bf E} - \theta \left[ {\ddot {\bf E} - \partial _0 \left( {\nabla
\times {\bf B}} \right) - 2\nabla \left( {\nabla  \cdot {\bf E}} \right)} \right]. \label{A17}
\end{equation}
Again, it is easy to see that in the electrostatic case ($\dot{\bf E}=0$ and
${\bf B}=0$), these equations reduce to
\begin{equation}
{\bf \Pi} ^{\left( 1 \right)}  = 0, \label{A18}
\end{equation}
and
\begin{equation}
{\bf p} =  - {\bf E} + 2\theta \nabla \left( {\nabla\cdot {\bf E}} \right).
\label{A19}
\end{equation}
Also, it should be noted that $p^0  = 0$ and $\Pi _0^{\left( 1 \right)}  = 0$.

Effectively, therefore, in the electrostatic case our canonical Hamiltonian takes
the form
\begin{equation}
H_C  = \int {d^3 x\left( {p_i \dot A^i  - {\cal L}} \right)}, \label{A20}
\end{equation}
which is the standard Legendre transformation. Thus the agreement between the
result (\ref{A9}) and our previous Hamiltonian treatment \cite{Accioly:2010js}
has been clarified. It follows from this that although the Lagrangian density
(\ref{A3}) contains higher time derivatives, in the electrostatic case  the
canonical momentum conjugate to velocities disappear. Therefore, the new Legendre
transformation reduces to the standard Legendre transformation.

Evidently, the next step would be to verify, order by order, that the above
conclusion is preserved. The above result suggests that this is the case.
On the other hand, and in order to support this observation we also call
attention to the fact that our physical state is defined solely in terms of
$A_{i}$-fields, that is,
\begin{eqnarray}
\left| \Phi  \right\rangle  &\equiv& \left| {\overline \Psi  \left(
\bf y \right)\Psi \left( {\bf y}\prime \right)} \right\rangle 
= \overline \psi \left( \bf y \right)\exp \left(
{iq\int\limits_{{\bf y}\prime}^{\bf y} {dz^i } A_i \left( z \right)}
\right)\psi \left({\bf y}\prime \right)\left| 0 \right\rangle.
\label{A21}
\end{eqnarray}
One can now observe that, although in principle there may be canonical momentum
conjugate to velocities, just will contribute to the calculation of the energy
only terms proportional to the canonical momentum conjugated to $A_{i}$. This
is consistent because our physical state is a gauge-invariant state, so this
state can not contain terms proportional to the velocities in order to preserve
gauge invariance. In other terms, our physical state is essentially a static
state.

\section{Appendix B: Non-perturbative vs perturbative results.}

In this appendix we would like to show the difference between
exact calculations and perturbative expansion in $\theta$. We shall
consider the simple case of the classical Coulomb potential in a 
non-commutative (euclidean) background geometry.
 The Poisson equation reads

\begin{equation}
\nabla^2 \phi\left(\,\vec{x}\,\right)=
-4\pi\, e\, \exp{\left(\,\theta \nabla^2\,\right)}
\delta\left(\,\vec{x}\,\right)\label{poisson}
\end{equation}
where the r.h.s. of equation (\ref{poisson}) is the smeared source
obtained through the rule (\ref{due}).
The Poisson equation can be \textit{exactly} solved through
standard Fourier method. The resulting Coulomb potential is

\begin{equation}
\phi\left(\,\vec{x}\,\right)= -\frac{e}{\sqrt\pi \, |\vec{x}|}
\gamma\left(\, 1/2\ ;\vec{x}^{\, 2}/4\theta\,\right)
\end{equation}
$\phi\left(\,\vec{x}\,\right)$ is regular in the origin:
\begin{equation}
\phi\left(\,0\,\right)= -\frac{e}{\sqrt\pi \, \theta}
\end{equation}
The divergence of the classical Coulomb potential has been cured
by the short-distance fluctuations of the coordinates.

Suppose we ignore the possibility to obtain an exact solution
of equation (\ref{poisson}) and proceed through a ``perturbative''
approach, i.e. we expand the exponential operators in the r.h.s.
up to the first order in $\theta$. This is the standard procedure
adopted in hundreds of papers implementing non-commutative effects
through the Wigner-Weyl-Moyal $\ast$-product. In our toy-model this
is equivalent to approximate the Poisson equation with

\begin{equation}
\nabla^2\, \phi\left(\,\vec{x}\,\right)=
-4\pi\, e\,\left(\, 1 + \theta \nabla^2+O\left(\, \theta^2\,\right)\right)
\delta\left(\,\vec{x}\,\right)
\end{equation}

It is immediate to find

\begin{equation}
\phi\left(\,\vec{x}\,\right)= -\frac{e}{ |\vec{x}|}-
4\pi\, e\,\theta \delta\left(\,\vec{x}\,\right)+O\left(\, \theta^2\,\right)
\end{equation}

Thus, the order-$\theta$ result is the standard Coulomb potential plus
a \textit{divergent} contact interaction in $\vec{x}=0$.
It is easy to see that also  the successive corrections do not improve
the short distance behavior of $\phi$. This follows from the fact that
at any finite order in $\theta$ the charge keeps its \textit{point-like}
nature. Removing the divergence in  $\vec{x}=0$ is a \textit{non-perturbative}
effect which cannot be seen at any finite order in a $\theta$-expansion.
This is a key feature of non-commutative field theories which does not
seem to have been fully understood by people working in this research field.

\ack

One of us (PG) wants to thank the Physics Department of the
Universit\`a di Trieste and the Field Theory Group of the CBPF
for hospitality. This work was supported in part by Fondecyt (Chile)
grant 1080260 (PG).\\


\begin{thebibliography}{99}
\bibitem{Witten:1985cc}
  E.~Witten,
  Nucl.\ Phys.\  B {\bf 268}, 253 (1986).

\bibitem{Seiberg:1999vs}
  N.~Seiberg and E.~Witten,
  JHEP {\bf 9909}, 032 (1999)


\bibitem{Douglas:2001ba}
  M.~R.~Douglas, N.~A.~Nekrasov,
  Rev.\ Mod.\ Phys.\  {\bf 73}, 977-1029 (2001).


\bibitem{Szabo:2001kg}
  R.~J.~Szabo,
  Phys.\ Rept.\  {\bf 378}, 207-299 (2003)


\bibitem{Gomis:2000sp}
  J.~Gomis, K.~Kamimura and T.~Mateos,
  JHEP {\bf 0103}, 010 (2001)

\bibitem{Bichl:2001nf}
  A.~A.~Bichl, J.~M.~Grimstrup, L.~Popp, M.~Schweda and R.~Wulkenhaar,
  Int.\ J.\ Mod.\ Phys.\  A {\bf 17}, 2219 (2002)

\bibitem{Snyder} H. Snyder, Phys. Rev. {\bf 71}, 38 (1947).


\bibitem{Minwalla:1999px}
  S.~Minwalla, M.~Van Raamsdonk and N.~Seiberg,
  JHEP {\bf 0002}, 020 (2000)


\bibitem{Gomis:2000zz}
  J.~Gomis and T.~Mehen,
  Nucl.\ Phys.\  B {\bf 591}, 265 (2000)


\bibitem{Bahns:2002vm}
  D.~Bahns, S.~Doplicher, K.~Fredenhagen and G.~Piacitelli,
  Phys.\ Lett.\  B {\bf 533}, 178 (2002)

\bibitem{Smailagic:2003rp}
  A.~Smailagic and E.~Spallucci,
  J.\ Phys.\ A  {\bf 36}, L517 (2003)

\bibitem{Smailagic:2003yb}
  A.~Smailagic and E.~Spallucci,
  J.\ Phys.\ A  {\bf 36}, L467 (2003)

\bibitem{Smailagic:2004yy}
  A.~Smailagic and E.~Spallucci,
  J.\ Phys.\ A  {\bf 37}, 1 (2004)
  [Erratum-ibid.\  A {\bf 37}, 7169 (2004)]


\bibitem{Galluccio:2008wk}
  S.~Galluccio, F.~Lizzi and P.~Vitale,
  Phys.\ Rev.\  D {\bf 78}, 085007 (2008)
\bibitem{Galluccio:2009ss}
  S.~Galluccio, F.~Lizzi and P.~Vitale,
  JHEP {\bf 0909}, 054 (2009)

\bibitem{Banerjee:2009xx}
  R.~Banerjee, S.~Gangopadhyay and S.~K.~Modak,
  Phys.\ Lett.\  B {\bf 686}, 181 (2010)

\bibitem{Gangopadhyay:2010zm}
  S.~Gangopadhyay and D.~Roychowdhury,
  ``Voros product, noncommutative inspired Reissner-Nordstr{\'o}m black hole
  and corrected area law,''
  arXiv:1012.4611 [hep-th].

\bibitem{Basu:2011kh}
  P.~Basu, B.~Chakraborty and F.~G.~Scholtz,
  J.\ Phys.\ A A {\bf 44} (2011) 285204
  

\bibitem{Hammou:2001cc}
  A.~B.~Hammou, M.~Lagraa and M.~M.~Sheikh-Jabbari,
  Phys.\ Rev.\  D {\bf 66}, 025025 (2002)

\bibitem{Gaete:2004ga}
  P.~Gaete and E.~I.~Guendelman,
  Mod.\ Phys.\ Lett.\  A {\bf 20}, 319 (2005)

\bibitem{Gaete:2007ax}
  P.~Gaete and I.~Schmidt,
  Phys.\ Rev.\  D {\bf 76}, 027702 (2007)


\bibitem{Gaete:2007zn}
   P.~Gaete, E.~Spallucci,
  Phys.\ Rev.\  {\bf D77}, 027702 (2008).


\bibitem{Nicolini:2005de}
  P.~Nicolini, A.~Smailagic and E.~Spallucci,
  ``The fate of radiating black holes in noncommutative geometry,''
  arXiv:hep-th/0507226.

\bibitem{Nicolini:2005vd}
  P.~Nicolini, A.~Smailagic and E.~Spallucci,
  Phys.\ Lett.\  B {\bf 632}, 547 (2006)
  .

\bibitem{Spallucci:2006zj}
  E.~Spallucci, A.~Smailagic, P.~Nicolini,
  Phys.\ Rev.\  {\bf D73}, 084004 (2006)

\bibitem{Ansoldi:2006vg}
  S.~Ansoldi, P.~Nicolini, A.~Smailagic {\it et al.},
  Phys.\ Lett.\  {\bf B645}, 261-266 (2007).


\bibitem{Spallucci:2008ez}
  E.~Spallucci, A.~Smailagic, P.~Nicolini,
  Phys.\ Lett.\  {\bf B670}, 449-454 (2009).

\bibitem{Nicolini:2009gw}
  P.~Nicolini, E.~Spallucci,
  Class.\ Quant.\ Grav.\  {\bf 27}, 015010 (2010).

\bibitem{Smailagic:2010nv}
  A.~Smailagic, E.~Spallucci,
  Phys.\ Lett.\  {\bf B688}, 82-87 (2010).


\bibitem{Spallucci:2011rn}
  E.~Spallucci and S.~Ansoldi,
  Phys.\ Lett.\ B {\bf 701} (2011) 471

\bibitem{Rinaldi:2010zu}
  M.~Rinaldi,
  Mod.\ Phys.\ Lett.\  A {\bf 25}, 2805 (2010)

\bibitem{Nicolini:2009dr}
  P.~Nicolini and M.~Rinaldi,
  Phys.\ Lett.\  B {\bf 695}, 303 (2011)
  


\bibitem{Rinaldi:2009ba}
  M.~Rinaldi,
  Class.\ Quant.\ Grav.\  {\bf 28} (2011) 105022
  [arXiv:0908.1949 [gr-qc]].

\bibitem{Moeller:2002vx}
  N.~Moeller and B.~Zwiebach,
  JHEP {\bf 0210} (2002) 034

\bibitem{Zee} See, for example, A. Zee, {\it Quantum Field Theory in a
Nutshell} (Princeton University Press, Princeton, NJ, 2003). 


\bibitem{Accioly:2011zz}
  A.~Accioly, P.~Gaete, J.~Helayel-Neto, E.~Scatena and R.~Turcati,
  Mod.\ Phys.\ Lett.\ A {\bf 26} (2011) 1985.

\bibitem{Accioly:2010js}
  A.~Accioly, P.~Gaete, J.~Helayel-Neto,  R. Turcati, E. Scatena
  ``Exploring Lee-Wick finite electrodynamics,''
  [arXiv:1012.1045 [hep-th]].

\bibitem{Gaete:1997eg}
  P.~Gaete,
  Z.\ Phys.\  C {\bf 76}, 355 (1997).

\bibitem{Dirac} P. A. M. Dirac, Can. J. Phys. {\bf 33}, 650 (1955).

\bibitem{Lee-Wick} T. Lee and G. Wick, Nucl. Phys. {\bf B9}, 209 (1969);
Phys. Rev. {\bf D2}, 1033 (1079).

\bibitem{Podolsky} B. Podolsky, Phys. Rev. {\bf 62}, 68 (1942); B.
Podolsky and C. Kikuchi, Phys. Rev. {\bf 65}, 228 (1944); B. Podolsky
and P. Schwed, Rev. Mod. Phys. {\bf 20}, 40 (1948).

\bibitem{Ostrogradsky} M. Ostrogradsky, Mem. Acad. St. Pet. VI, {\bf 4},
385 (1850).

\bibitem{GalPim}  C. A. P. Galv\~ao and B. M. Pimentel, Can. J. Phys.
{\bf 66}, 460 (1988).
\end{thebibliography}
\end{document}